\documentclass[journal]{IEEEtran}
\IEEEoverridecommandlockouts

\usepackage{amsmath,amssymb,amsfonts}
\usepackage{graphicx, bbm}
\usepackage{textcomp, nccmath}
\usepackage{xcolor}
\usepackage{amsmath}
\usepackage{algorithm}
\usepackage{algorithmicx,algpseudocode}
\usepackage{subcaption}
\def\BibTeX{{\rm B\kern-.05em{\sc i\kern-.025em b}\kern-.08em
    T\kern-.1667em\lower.7ex\hbox{E}\kern-.125emX}}
\begin{document}
\title {Mobile Network Slicing under Demand Uncertainty: A Stochastic Programming Approach}

\author{\IEEEauthorblockN{ Anousheh Gholami, Nariman Torkzaban, and John S. Baras\\}
\IEEEauthorblockA{\textit{Department of Electrical and Computer Engineering,\\}{\& Institute for Systems Research,\\}
\textit{University of Maryland, College Park, Maryland, USA}\\
Email: \{anousheh, narimant, baras\}@umd.edu}
}

\maketitle

\begin{abstract}

Network slicing enables the deployment of multiple dedicated virtual sub-networks, i.e. slices on a shared physical infrastructure. Unlike traditional one-size-fits-all resource provisioning schemes, each network slice (NS) in 5G is tailored to the specific service requirements of a group of customers. An end-to-end (E2E) mobile NS orchestration requires the simultaneous provisioning of computing, storage, and networking resources across the core network (CN) and the radio access network (RAN). Constant temporospatial changes in mobile user demand profiles further complicate the E2E NSs resource provisioning beyond the limits of the existing best-effort schemes that are only effective under accurate demand forecasts for all slices. 

This paper proposes a practical two-time-scale resource provisioning framework for E2E network slicing under demand uncertainty. At each macro-scale instance, we assume that only the spatial probability distribution of the NS demands is available. We formulate the NSs resource allocation problem as a stochastic mixed integer program (SMIP) with the objective of minimizing the total resource cost at the CN and the RAN. At each micro-scale instance, utilizing the exact slice demand profiles, a linear program is solved to jointly minimize the unsupported traffic and the resource cost at the RAN. We verify the effectiveness of our resource allocation scheme through numerical experiments.

\end{abstract}

\section{Introduction}
Softwarization based on network function virtualization (NFV) and software-defined networking (SDN) is envisioned to realize programmable control and management of network resources in next-generation mobile networks \cite{barakabitze20205g}. Building upon SDN and NFV as the enabling technologies, \emph{network slicing} allows the physical infrastructure to be segmented into several logically-isolated sub-networks each customized to the specific quality of service (QoS) requirements of an application. Such a dedicated resource provisioning framework results in significant reductions in network management and equipment costs \cite{wireless-virtu}.

In the paradigm of 5G enhanced by \emph{network slicing}, mobile network operators (MNOs) manage and set up network slices (NSs) and provide service providers (SPs) with an on-demand scalable delivery of network services \cite{5g-NS}. The SPs, a.k.a. \textit{tenants}, dedicate NSs to customers with various QoS requirements in a sustainable manner. A mobile NS spans across multiple domains, namely the radio access network (RAN), the core network (CN), and the transport network (TN), forming an end-to-end (E2E) sub-network. 

In contrast to the traditional static allocation of resources \cite{RL-NS}, 5G RAN slicing introduces the capability of sharing physical network infrastructure among mobile virtual network operators (MVNOs). Consequently, the radio resources such as time, frequency, and power are reserved on the fly based on dynamic user demands. Each sub-slice corresponding to a specific part of the network includes service function chains (SFCs) consisting of multiple virtual network functions (VNFs) and physical network functions (PNFs). Designing E2E NSs requires resource provisioning across heterogeneous physical and virtual network infrastructures each having specific technical constraints. On the other hand, with a proliferation of advanced and diverse digital services such as hologram video streaming and multi-sensory extended reality, the sixth-generation mobile system (6G) is envisioned to facilitate massive and extremely heterogeneous \emph{network slicing} where tenancy would be extended further to the end user \cite{6G-slice}. 
In such architectures with slicing-aware user equipment (UE), new challenges are introduced. For instance, UE mobility should be considered by the SPs as part of the slice setup and management procedure \cite{ns-scal-flex}.

Although \emph{network slicing} promises to enhance the agility of next-generation mobile networks, practical NS deployment faces key challenges among which the uncertainty in the demand for different slices stands out. In practice, the shared physical network resources must be dynamically and efficiently allocated to logical NSs based on changing user demands. However, the existing resource provisioning methods for \emph{network slicing} are typically performed in a best-effort manner \cite{su2019resource}. Consequently, there is no guarantee that resources allocated to different slices will be sufficient to support constantly-varying demand profiles. Besides the dynamicity of the user demand profiles, the variation in the infrastructure resource availability status may degrade the slice QoS compared to the service level agreement (SLA) promised by the SPs \cite{luu2021uncertainty}. \emph{Elasticity} which refers to the availability of resources in an adaptive way according to user demands plays a critical role in meeting the differentiated QoS requirements of different slices while avoiding resource under-utilization or over-utilization \cite{khan2020network}. 

In this paper, we propose a novel approach to optimize the E2E resource provisioning for \emph{network slicing} under demand uncertainty. Stochastic programming is a powerful tool to address optimization under uncertainty. We consider the joint resource allocation of next-generation RAN (NG-RAN) and 5G core (5GC) for different slices as shown in Fig. \ref{fig:E2ENS}. 
In our proposed solution, the RAN slicing is triggered more frequently (i.e. has a shorter life cycle) compared to the CN segment, due to the existence of more dynamic parameters in the wireless network such as user mobility and channel conditions. Therefore, our algorithm operates at two time scales. At each macro-scale instance, we assume that only the spatial probability distribution of the network slice demands is available. We formulate the network slices resource allocation problem as a stochastic mixed integer program (SMIP) with the objective of minimizing the total resource cost at the CN and the RAN. At each micro-scale instance,  utilizing the exact slice demand profiles, a linear program is solved to jointly minimize the unsupported traffic and the resource provisioning cost at the RAN.


The remainder of the paper is organized as follows. In section~\ref{sec:related}, we present the literature review.  Section~\ref{sec:sys} describes the system model. The problem formulation is provided in section~\ref{sec:formulation_thesis}. The proposed solution is presented in section~\ref{sec:solution}, while the numerical results are provided in section \ref{sec:evaluation}. Finally, we highlight our conclusions in section \ref{sec:conclusion}.

\section{Related Work}
\label{sec:related}
 To realize \emph{network slicing} in next-generation mobile networks, different required mechanisms including resource provisioning (reservation), slice admission control, resource scheduling, and resource orchestration \cite{hurtado2022deep} have been investigated in the literature for the RAN, TN, CN, or E2E domains.
Authors in \cite{9133773} develop an admission control scheme for sliced 5G networks considering inter-slice and intra-slice priorities in order to enhance UEs' QoE and network resource utilization. In \cite{E2E-raof}, the E2E (RAN and CN) resource orchestration of 5G NSs is addressed and a heuristic algorithm for the slice placement problem with the objective of minimizing the number of VNF migrations in the network while optimizing network utilization and satisfying QoS is proposed. 

Authors in \cite{xiang2019joint} study the resource allocation problem of 5G networks leveraging both \emph{network slicing} and mobile edge computing (MEC). They propose an optimization problem with the objective of minimizing the total latency of transmitting, outsourcing and processing user traffic, under the constraint of tolerable latency. In \cite{Tok}, an auction-based model for the joint resource and revenue optimization of RAN and CN NS segments. In all of the aforementioned works, it is assumed that the demand for different slices is known in advance which is an unrealistic assumption. 

In practice, the task of accurately estimating future traffic behavior is very challenging \cite{alliance20155g}. To alleviate this problem, authors in \cite{gholamipour2021online} propose a solution for the joint online admission control and resource allocation problem of \emph{network slicing} using the $\Gamma$-Robustness concept to consider the uncertainty in computation and communication demands. The main drawback of this method is that they only consider the CN slicing problem and ignore the correlation between the demands for different slices. Similarly, a robust model is proposed in \cite{baumgartner2018towards} for the NS design problem under demand uncertainty. In this work, the authors consider both uncorrelated and correlated traffic case which resembles the spatial demand correlations, but they only study the CN design problem. Moreover, they do not consider delay constraints and the differentiation between different slices based on delay requirements. 






\section{System Model}
\label{sec:sys}

\begin{figure}[!t]
    \centering
    \includegraphics[width=0.4888\textwidth]{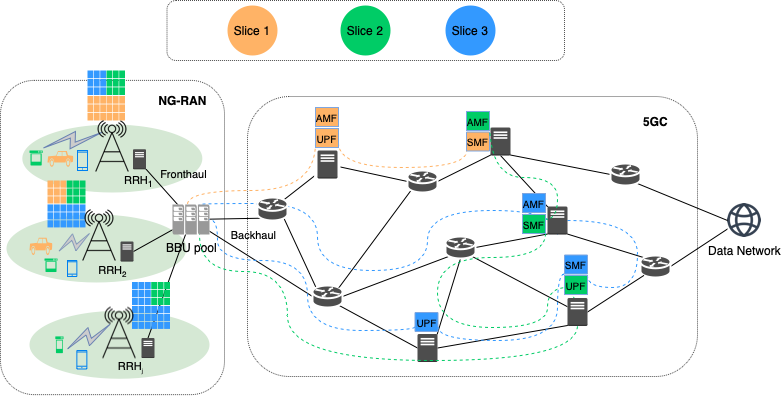}
    \caption{E2E resource provisioning for multiple NSs}
    \label{fig:E2ENS}
\end{figure}


\subsection{Substrate Network Model}
We consider a mobile network infrastructure (a.k.a. substrate network) that is comprised of next-generation nodeBs (gNBs) and CN nodes implementing 5GC components. In a mobile network empowered by NFV and SDN, the substrate nodes are general-purpose servers that can host various VNFs. Fig.~\ref{fig:E2ENS} illustrates such an infrastructure. Let $G=(V,E)$ denote the substrate network graph, where $V$ represents the set of substrate nodes and $E$ denotes the set of infrastructure links. We assume that $V = V_{gNB} \cup V_{-gNB}$ where $V_{gNB}$ and $V_{-gNB}$ are the substrate gNB and non-gNB nodes, respectively. Each gNB is characterized by its maximum supported traffic that is computed based on its available resources, or physical resource blocks (PRBs), and antenna configuration. Non-gNB nodes are general-purpose servers providing essential capabilities for running core VNFs. A substrate node is characterized by its residual CPU, storage, and RAM resources. The vector $W_j = (W^{CPU}_j, W^{STO}_j, W^{RAM}_j)$ represents the residual capacity of the substrate node $j\in V$, where we define the set of resources as $\mathcal{T} = \{CPU, STO, RAM\}$. 

Similar to the RAN slicing model considered by \cite{papa2021user}, we assume that the gNB $j\in V_{gNB}$ has $W^{r}_j$ available PRBs that can be allocated to the users of different slices. Furthermore, we assume that $E = E_{FH} \cup E_{BH}$ where $E_{FH}$ and $E_{BH}$ stand for the set of gNB-CN links and the remaining links, respectively. Each substrate link $e\in E$ is characterized by its bandwidth capacity ($W^{BW}_e$) and propagation delay ($\tau^{PRO}_e$). Moreover, let $\mathcal{P}$ denote the set of substrate paths used for traffic routing. Define $\mathcal{P}( i \rightarrow j )$ to be the set of substrate paths between nodes $i$ and $j$. Therefore, $\mathcal{P} = \cup_{i,j\in V, i\neq j} \mathcal{P}( i \rightarrow j )$.
Let $\mathcal{U}$ denote the set of UEs distributed across the geographical area under cover. We assume that each UE is served by a gNB according to the nearest association rule, whereby a UE is assigned to the gNB achieving the maximum SNR.

\subsection{Network Slice Model}
We assume that an NS contains one or multiple SFCs, each comprised of a number of VNFs (e.g. gNB, AFM, UPF, SMF) and virtual links (VLs) between them. The set of NSs is denoted by $\mathcal{K}=\{1,...,K\}$. Let $G^\prime_{k} = (V^\prime_{k}, E^\prime_{k})$ represent the $k$th slice SFC modeled as an undirected graph. 
We assume $V^\prime_{k} = V^\prime_{k,gNB} \cup V^\prime_{k,-gNB}$ where $V^\prime_{k,gNB}$ and $V^\prime_{k,-gNB}$ correspond to the set of gNBs and non-gNB VNFs of the $k$th slice SFC. Moreover, $E^\prime_{k} =  E^\prime_{k, FH}\cup E^\prime_{k, BH}$ where $E^\prime_{k, FH}$ and $E^\prime_{k, BH}$ denote the fronthaul and backhaul VLs, respectively. We assume that each instance of the SFC corresponding to slice $k$ has QoS requirements expressed as network-level UE throughput and maximum E2E tolerable latency. In order to guarantee the performance requirement of different slices, the network-level performance metric of a slice is translated to cell-level radio resource requirement \cite{8624268}. Let $\underline{R}_j^k$ denote the average required number of PRBs for a UE requesting slice $k\in \mathcal{K}$ covered by gNB $j\in V_{gNB}$. The value of $\underline{R}_j^k$ depends on the overall cell load, antenna configuration, channel condition, modulation and coding scheme (MCS), and slice performance requirements. The details of such a translation mechanism are beyond the scope of this article. We encourage interested readers to refer to \cite{shi2021two} and \cite{oliveira2022mapping} for prospective solutions.

We assume that the traffic flow of each $e^\prime \in E^\prime_k$ is routed through substrate paths. Therefore, we represent the set of substrate paths considered for slice $k$ edges by $\mathcal{P}_k$, where $\mathcal{P}_k \subseteq \mathcal{P}$. Let $\mathcal{P}^\prime_k$ denote the set of all paths of $G^\prime_k$. Each path of $G^\prime_k$ corresponds to either CP or UP data flows. For instance, the path gNB-AMF-SMF of a 5G NS is a CP path while gNB-UPF is a UP path. Let $\mathcal{P}^\prime_{k,UP}$ and $\mathcal{P}^\prime_{k,CP}$ denote the set of UP and CP paths in $G^\prime_k$. 
We assume that the QoS requirement of each NS is given as the maximum tolerable UP and CP latencies denoted by $d_{k}^{UP},d_{k}^{CP}$, respectively. Let $\mathcal{U}_k \subseteq \mathcal{U}$ denote the set of UEs requesting slice $k\in\mathcal{K}$. We define $u_j^{k,i}$ to be equal to $1$ if user $i$ of slice $k$ ($i\in \mathcal{U}_k$) is covered by gNB $j\in V_{gNB}$ and $0$, otherwise. 

In order to support the demands of all users requesting an NS, multiple instances of the slice may need to be deployed. Let $I_k$ be the number of instances of slice $k$ deployed to support the load for slice $k$. In addition, each instance of the VNF $j^\prime \in V^\prime_k$ requires CPU, storage, and RAM resources. The required amount of these resources depends on the slice load and the resource-sharing factor which determines how much of the workload of a VNF can be shared among multiple UEs. Let the vector $R_{j^\prime}= (R^{CPU}_{j^\prime}, R^{STO}_{j^\prime}, R^{RAM}_{j^\prime})$ denote the required per-unit amount of CPU, STO and RAM resources for VNF $j^\prime$. Similarly, each VL $e^\prime \in E^\prime_{k}$ is characterized by its per-unit bandwidth requirement $R^{BW}_{e^\prime}$ to meet the demand for data transmission between the two end-points VNFs of $e^\prime$. The per-unit resource requirements must be scaled according to the traffic load and resource-sharing factor. We define $\chi^t_{k,j^\prime}$ to be the scaling factor of resource $\nu$, $\nu\in \mathcal{T}$ for VNFs of slice $k$.

Moreover, $\chi^{BW}_{k,e^\prime}$ stands for the BW scaling factor of VLs of slice $k$. 
The cost of running a VNF on the infrastructure node $j$ is composed of two parts: (i) a fixed cost denoted by $C_j$, (ii) a variable cost that increases linearly with respect to the consumed amount of resources by that VNF. Let $C_j^{CPU}, C_j^{STO} $, and $C_j^{RAM}$ denote the per-unit cost of using the CPU, storage, and memory of node $j$, respectively. The per-unit bandwidth cost of the infrastructure link $e$ is represented by $C_e^{BW}$. 
For the sake of simplicity and without loss of generality, we assume that each UE requests at most one slice. A summary of the notations used in this paper is provided in Table \ref{chapter2:parameters}.


\begin{table}[t]
\centering
\caption{System Model Parameters}
\begin{center}
\scalebox{0.93}{
\begin{tabular}{c|c}
\hline
\hline
Network Parameters & Description\\
\hline
$G = (V, E)$ & Substrate network graph\\
$V_{gNB}, V_{-gNB}$ & Set of substrate gNB and non-gNB nodes\\
$E_{FH}, E_{BH}$ & Set of fronthaul and backhaul links\\
$\mathcal{P}_k$ & Set of substrate paths considered for slice $k$\\
$W_{j}^{r}$ & Available spectrum of gNB $j$\\
$\mathcal{U}_k, U_k$ & Set and number of users of slice $k$\\
$W_j^{CPU,STO,RAM}$ & CPU, Storage and RAM capacities of node $j$\\
$W_e^{BW}$ & BW capacity of the substrate link $e$\\
\hline
\hline
Slice Parameters & Description\\
\hline
$\mathcal{K}, K$ & Set and number of requested slices\\
$G^\prime_{k} = (V_{k}^\prime, E_{k}^\prime)$& The $k$th slice SFC graph\\
$V^\prime_{k,gNB}, V^\prime_{k,-gNB}$ & Set of virtual RAN and core VNFs\\
$E_{k, FH}^\prime, E_{k, BH}^\prime$ & Set of fronthaul and backhaul virtual links\\
$\mathcal{P}^\prime_k$ & Set of slice $k$ E2E paths\\
$\overline{R}_j^k$ & Slice $k$ maximum allowed radio resource in gNB $j$\\
$\underline{R}_j^k$ & Slice $k$ radio resource requirement in gNB $j$\\
$R_{j^\prime}^{CPU, STO, RAM}$ & CPU, Storage and RAM requirements of the VNF $j^{\prime}$\\
$R_{e^\prime}^{BW}$ & BW requirement of the virtual link $e^\prime$\\
$d_{k}^{UP}, d_{k}^{CP}$ & UP and CP latency requirement of slice $k$\\
\hline
\hline
\end{tabular}}
\end{center}
\label{chapter2:parameters}
\end{table}


\section{Problem Formulation}
\label{sec:formulation_thesis}
In this section, we formulate the E2E network slicing problem under demand uncertainty as a two-stage SMIP. Before we present the problem formulation, in the next subsection we provide a brief overview of stochastic programming \cite{book}. 

\subsection{Stochastic Programming}
\label{sec:prelim}
\textit{Stochastic linear programs} (SLPs) are linear programs where some data are uncertain, represented as random variables with given probability measures. The values of these random variables are only known after a realization of a random experiment. SLP optimization decision variables are classified into two groups: first-stage variables, determined before conducting the experiment, and second-stage variables, determined after the experiment outcome is observed.
Let $\boldsymbol{x} \in \mathbb{R}^{n_1\times 1}$ denote the vector of first-stage optimization variables. Define the random variable corresponding to the unknown data as $\boldsymbol{\xi}=\boldsymbol{\xi}(\omega)$ with $\omega\in \Omega$ representing a random outcome. The vector of second-stage decision variables is denoted by $\boldsymbol{y}(\omega,\boldsymbol{x})$. We assume that the probability distribution $\mathcal{F}$ on $\Omega$ is known in the first stage. The classical two-stage SLP with \textit{fixed recourse} is formulated as follows:
\begin{align}
    & minimize \quad \boldsymbol{c}^T \boldsymbol{x} + \mathbb{E}_{\mathcal{F}}[min\ \boldsymbol{q}(\omega)^T \boldsymbol{y}(\omega,\boldsymbol{x})] \label{SLP-obj}\\
    & s.t. \quad A\boldsymbol{x} = \boldsymbol{b}\label{SLP-c1}\\
    & T(\omega)\boldsymbol{x} + W \boldsymbol{y}(\omega,\boldsymbol{x}) = \boldsymbol{h}(\omega)\label{SLP-c2}\\
    & \boldsymbol{x},\boldsymbol{y}(\omega,\boldsymbol{x}) \geq \boldsymbol{0} \label{SLP-c3}.
\end{align}
where $\boldsymbol{c}\in \mathbb{R}^{n_1\time 1}, \boldsymbol{b}\in \mathbb{R}^{m_1\times 1}, A\in \mathbb{R}^{m_1\times n_1}$. The objective function \eqref{SLP-obj} consists of the deterministic term $\boldsymbol{c}^T \boldsymbol{x}$ and the expectation of the term $\boldsymbol{q}(\omega)^T \boldsymbol{y}(\omega,\boldsymbol{x})$ over all realizations $\omega$. Note that once $\omega$ is realized, $T(\omega) \in \mathbb{R}^{m_2\times n_2}, \boldsymbol{h}(\omega) \in \mathbb{R}^{m_2\times 1}$, and $\boldsymbol{q}(\omega)\in \mathbb{R}^{n_2\times 1}$ become known. In particular $\boldsymbol{\xi}(\omega) = (\boldsymbol{q}(\omega), \boldsymbol{h}(\omega), T_1(\omega),\dots,T_{m_2}(\omega))$ represents the random data vector where $T_1(\omega),\dots,T_{m_2}(\omega)$ are the rows of $T(\omega)$. For a given $\omega$, let $Q(\boldsymbol{x},\boldsymbol{\xi}(\omega))$ denote the value of the second-stage problem, defined as:
{
\begin{align}
    & Q(\boldsymbol{x},\boldsymbol{\xi}(\omega)) = min_{\boldsymbol{y}\geq 0} \{\boldsymbol{q}(\omega)^T \boldsymbol{y} | W\boldsymbol{y} = \boldsymbol{h}(\omega)-T(\omega)\boldsymbol{x}\} \label{2-stage-value}
\end{align}
}

Moreover, the expected value of the second-stage objective value is defined as:
\begin{align}
    & \mathcal{Q}(\boldsymbol{x}) = \mathbb{E}_{\mathcal{F}}[Q(\boldsymbol{x},\boldsymbol{\xi}(\omega))] \label{2-stage-expected-value}
\end{align}
The \textit{deterministic equivalent program} (DEP) corresponding to \eqref{SLP-obj}-\eqref{SLP-c3} is defined as:
\begin{align}
    & min \quad \boldsymbol{c}^T \boldsymbol{x} + \mathcal{Q}(\boldsymbol{x}) \label{DEP-obj}\\
    & s.t. \quad A\boldsymbol{x} = \boldsymbol{b} \label{DEP-c1}\\
    & \boldsymbol{x} \geq \boldsymbol{0} \label{DEP-c2}
\end{align}
The DEP formulation illustrates the major difference between a deterministic linear program and an SLP. Given $\mathcal{Q}(\boldsymbol{x})$, an SLP is an ordinary nonlinear program due to the nonlinearity of $\mathcal{Q}(\boldsymbol{x})$. On the other hand, the computation of $\mathcal{Q}(\boldsymbol{x})$ requires evaluating $Q(\boldsymbol{x},\boldsymbol{\xi}(\omega))$ for all $\boldsymbol{x}$ and $\omega$ that is not possible in practice. The problem becomes even more challenging considering the first-stage or second-stage variables to be integers, i.e. $\boldsymbol{x}\in \mathbb{Z}^{n_1}$, $\boldsymbol{y}\in \mathbb{Z}^{n_2}$. Next, we present our proposed formulation for the network slicing problem under demand uncertainty, which is based on two-stage stochastic mixed-integer programming.

\subsection{ Stochastic Mobile Network Slicing Optimization Model}
\label{sec:SMNS}
We start by defining the decision variables for the optimization model as follows:
\begin{itemize}
\item $\boldsymbol{x}_{k}$: the set of binary variables where $x_j^{k,j^\prime}$ is equal to $1$ if the VNF $j^\prime$ of the $k$th slice is placed on the substrate node $j$.
\item $\boldsymbol{y}_{k}$: the set of binary variables where $y_p^{k,e^\prime}$ is equal to $1$ if the VL $e^\prime$ of the $k$th slice is mapped to substrate path $p\in\mathcal{P}_k$.
\item $\boldsymbol{z}_k$: the set of continuous variables where $z_j^k$ represents the fraction of gNB $j$ spectrum allocated to slice $k$.
\end{itemize}
The vector $\boldsymbol{z} = (z_1^1,...,z_{|V_{gNB}|}^1,...,z_1^K,...,z_{|V_{gNB}|}^K)$ represents the RAN slicing policy defined as the fraction of residual radio resource at each cell that is assigned to each slice. For instance, in the case of having three NSs, the resulting RAN slice allocation for a gNB can be $20\%, 50\% $, and $30\%$. Once this spectrum assignment policy is determined, the actual radio resource allocation in terms of the scheduled PRBs for the users of each slice is obtained from a separate problem that is out of the scope of this paper. In this regard, different solutions have been proposed in the literature. For instance, authors in \cite{RAN} proposed a RAN resource allocation solution given the spectrum assignment policy, with the objective of minimizing the interference between different mobile virtual network operators. 

We model the resource provisioning problem for the requested slices as a two-stage SMIP. We refer to this problem as stochastic mobile network slicing (\emph{SMNS}). The two stages of the problem are explained in the following.

\subsubsection{First-Stage Problem}
The first-stage objective of \emph{SMNS} is the minimization of the total provisioning cost which consists of node and link deployment costs for all slices. Let $\mathcal{C}_k^{N}$ and $\mathcal{C}_k^{L}$ denote the node and link costs corresponding to slice $k$, respectively. Consequently, we have:
\begin{equation}
    \mathcal{C}_k^N(\boldsymbol{x}_{k}) = \sum_{j\in V} \sum_{j^{\prime}\in V^\prime_{k}} C_j^kx_{j}^{k}+\sum_{j\in V} \sum_{j^{\prime}\in V^\prime_{k}}\sum_{\nu\in \mathcal{T}}\gamma_\nu R_{j^\prime}^\nu C_j^\nu x_j^{k,j^\prime}
    \label{SMIP:node-cost}
\end{equation}
\begin{equation}
    \mathcal{C}_k^L(\boldsymbol{y}_{k}) = \sum_{p\in\mathcal{P}}\sum_{e\in p}\sum_{e^\prime \in E^\prime_k}\gamma_{BW} R_{e^\prime}^{BW} C_e^{BW} y_{p}^{k, e^\prime}
    \label{SMIP:link-cost}
\end{equation}
where
\begin{equation}
  x_{j}^{k} =
    \begin{cases}
      1 & \text{if $\sum_{j^\prime\in V^\prime_{k}}x_j^{k,j^\prime} \geq 0$}\\
      0 & \text{otherwise}
    \end{cases}       
\end{equation}

Hence, the first-stage objective is: \begin{equation}
     \mathcal{C}_1 (\boldsymbol{x},\boldsymbol{y}) = \sum_{k\in\mathcal{K}}\mathcal{C}^N_k(\boldsymbol{x}_k)+ \sum_{k\in\mathcal{K}}\mathcal{C}^L_k(\boldsymbol{y}_k)
     \label{SMIP:Obj1}
\end{equation}
where $\gamma_\nu, \nu\in\mathcal{T}\cup\{BW\}$ are the weights used to balance the objective terms of \eqref{SMIP:Obj1} corresponding to different resources.
In a valid network slicing solution, each gNB VNF is placed at substrate gNB nodes only and each non-gNB VNF runs on a substrate non-gNB node. Thus, the resulting VNFs deployment must satisfy the following constraints:
\begin{equation}
    \sum_{j\in V_{-gNB}}x_j^{k,j^\prime} = 1, \ \forall k \in \mathcal{K}, j^\prime \in V^\prime_{k,-gNB}
    \label{SIMP:non-gNB-map1}
\end{equation}
\begin{equation}
    x_j^{k,j^\prime} = 0, \ \forall k \in \mathcal{K},j\in V_{-gNB}, j^\prime \in V^\prime_{k,gNB}
    \label{SIMP:gNB-map1} 
\end{equation}
The flow conservation is guaranteed by constraints \eqref{SIMP:flow2-nongNB} and \eqref{SIMP:flow2-gNB}, 
\begin{align}
    & \sum_{\substack{{}p \in \mathcal{P}_k (i\rightarrow j)\\q \in \mathcal{P}_k (j\rightarrow i)\\ j\in V}} y_{p}^{k,e^\prime} -y_{q}^{k,e^\prime} = x_{i}^{k,i^\prime}-x_{i}^{k,j^\prime},\nonumber \\ 
    & \forall k\in\mathcal{K}, e^\prime \in E^\prime_{k,BH}, src(e^\prime) = i^\prime, dst(e^\prime) = j^\prime \label{SIMP:flow2-nongNB}
\end{align}
\begin{align}
    & \sum_{\substack{p \in \mathcal{P}_k(i\rightarrow j)\\ q\in \mathcal{P}_k(j\rightarrow i) \\ j\in V}} y_{p}^{k,e^\prime} - y_{q}^{k,e^\prime} = (x_{i}^{k,i^\prime}-x_{i}^{k,j^\prime})I_{k},\nonumber \\ 
    &\forall k\in \mathcal{K}, i\in V_{-gNB}, e^\prime\in E_{k,FH}^\prime, src(e^\prime) = i^\prime, dst(e^\prime) = j^\prime \label{SIMP:flow2-gNB} 
\end{align}
The domain constraints are as follows,
{\small
\begin{align}
    & x_j^{k,j^\prime}, y^{k,e^\prime}_p \in \{0,1\}  \ \forall k\in \mathcal{K}, j\in V, j^\prime \in V^\prime_k, p\in \mathcal{P}_k, e^\prime \in E^\prime_k \label{SMIP:dom1}
\end{align}
}
\subsubsection{Second-Stage Problem} The goal of the second-stage problem is to minimize the total cost of gNBs' radio resources allocated to different NSs. 
The second-stage objective is:
\begin{equation}
     \mathcal{C}_2(\boldsymbol{z}) = \sum_{k\in\mathcal{K}}\sum_{j\in V_{gNB}} C^k_j z^k_{j} W^r_j
\end{equation}

The processing delay of VNF $j^\prime \in V^\prime_k$ on an infrastructure node is represented by $\tau^{k,j^\prime}$. Given the VL mapping decision variables $y^{k,e^\prime}_p$, the data transmission latency corresponding to the infrastructure edge $e\in E$ given its aggregated load is calculated as:
\begin{equation}
    \sum_{k\in\mathcal{K}} \sum_{p\in\mathcal{P}_k:e\in p}\sum_{e^\prime \in E^\prime_k} y^{k,e^\prime}_p \frac{\chi_{k}^{BW}R^{BW}_{e^\prime}}{W^{BW}_e} \label{d-edge} \nonumber
\end{equation}
Thus, the latency of the VL $e^\prime \in E^\prime_k$ mapped to the path $p\in \mathcal{P}_k$ is:
\begin{equation}
    \sum_{e\in p} \left[\sum_{k\in\mathcal{K}} \sum_{p\in\mathcal{P}_k:e\in p}\sum_{e^\prime \in E^\prime_k} y^{k,e^\prime}_p \frac{\chi_{k}^{BW}R^{BW}_{e^\prime}}{W^{BW}_e}\right] \nonumber
\end{equation}
Therefore, the latency of VL $e''$ is calculated as follows:
\begin{equation}
    \sum_{p\in\mathcal{P}_k} y^{k,e''}_p \left[\sum_{e\in p} \left[\sum_{k\in\mathcal{K}} \sum_{p\in\mathcal{P}_k:e\in p}\sum_{e^\prime \in E^\prime_k} y^{k,e^\prime}_p \frac{\chi_{k}^{BW}R^{BW}_{e^\prime}}{W^{BW}_e}\right]\right] \label{d-slice-edge} \nonumber
\end{equation}
Thus, the maximum UP/CP latency of slice $k$ instances are guaranteed to be lower than the required latency $d^{UP}_k$/$d^{CP}_k$ in the following constraint:
{\small
\begin{align}
    & \sum_{e'' \in p^\prime} \left[ \sum_{p\in\mathcal{P}_k} y^{k,e''}_p \left[\sum_{e\in p} \left[\sum_{k\in\mathcal{K}} \sum_{p\in\mathcal{P}_k:e\in p}\sum_{e^\prime \in E^\prime_k} y^{k,e^\prime}_p \frac{\chi_{k}^{BW}R^{BW}_{e^\prime}}{W^{BW}_e}\right]\right] \right] \nonumber\\
    &+ \sum_{j^\prime \in p^\prime} \left[ \sum_{j\in V} x^{k,j^\prime}_j \tau^{k,j^\prime} \right]  \leq d_k^{UP/CP}, \forall p^\prime \in \mathcal{P}^{\prime}_{k,UP/CP}, k\in \mathcal{K} \nonumber
\end{align}}
The above constraint ensures that the maximum latency of each slice is less than its tolerable latency by enforcing the inequality for all paths of $G^\prime_k$. By doing so, the latency over the \textit{critical path} defined as the path inducing the maximum latency among all paths is enforced to be less than the tolerable latency. This constraint is a quadratic constraint. Thus, we linearize it by introducing a set of additional  continuous variables $\eta_p \geq 0$ and $\eta_p^{k,e^\prime} \geq 0$, defined as the latency of path $p\in\mathcal{P}$ and the latency of VL $e^\prime \in E^\prime_k$ of slice $k$ on path $p$, respectively. Therefore, the latency constraint is converted to the following set of constraints:
{
\begin{align}
    & \sum_{e\in q}\left[ \sum_{k\in\mathcal{K}} \sum_{p\in\mathcal{P}_k:e\in p} \sum_{e^\prime \in E^\prime_k} y^{k,e^\prime}_p\frac{\chi_{k}^{BW}R^{BW}_{e^\prime}}{W^{BW}_e}\right] = \eta_q, \ \forall q\in \mathcal{P} \label{SIMP:latency-1}
\end{align}
}
\begin{align}
    &\zeta y^{k,e^\prime}_p + \eta_p - \eta^{k,e^\prime}_p \leq \zeta, \ \forall p\in\mathcal{P}_k, e^\prime \in E^\prime_k, k\in\mathcal{K}\label{SIMP:latency-2}
\end{align}
\begin{align}
    &\sum_{e^\prime \in p^\prime} \sum_{p\in \mathcal{P}_k} \eta^{k,e^\prime}_p + \sum_{j^\prime \in p^\prime} \sum_{j\in V} x^{k,j^\prime}_j \tau^{k,j^\prime} \leq d^{UP/CP}_k,\nonumber \\ &\forall p^\prime \in \mathcal{P}^\prime_{k,UP/CP}, k\in \mathcal{K} \label{SIMP:latency-3}
\end{align}
where $\zeta$ is a large real constant. We also add the term $\epsilon \mathcal{D}$ to the objective function where $\mathcal{D}(\boldsymbol{\eta})=\sum_{k\in \mathcal{K}}\sum_{p\in \mathcal{P}_k}\sum_{e^\prime\in E^\prime_k}\eta_p^{k,e^\prime}$ and $\epsilon$ is a very small value in order to make sure that the main objective of minimizing the resource provisioning cost is not affected by adding $\mathcal{D}$. By doing so, it is ensured that the value of $\eta_p^{k,e^\prime}$ is positive if and only if $y_p^{k,e^\prime} = 1$. 

Given the network-level performance translation to cell-level metric $\underline{R}^k_j$, the RAN slice resource allocation is enforced by the following constraints:
\begin{equation}
     \sum_{i=1}^{U_k}u_{j}^{k,i}\underline{R}^k_j \leq x^{k,j^\prime}_j z^k_{j}W^r_j, \ \forall k\in\mathcal{K}, j\in V_{gNB}, j^\prime \in V_{k,gNB}^\prime \label{SIMP:reqrate}
\end{equation}
Constraints \eqref{SIMP:reqrate} is nonlinear due to the term $x^{k,j^\prime}_j z^k_{j}W^r_j$. We use the big-M method and convert \eqref{SIMP:reqrate} to the following:
\begin{align}
     & \sum_{i=1}^{U_k}u_{j}^{k,i}\underline{R}^k_j \leq z^k_{j}W^r_j+(1-x^{k,j^\prime}_j)M, \ \forall k\in\mathcal{K}, \nonumber\\
     & j\in V_{gNB}, j^\prime \in V_{k,gNB}^\prime \label{SIMP:reqrate-bigM}
\end{align}
where $M\in\mathbb{R}$ is a large number. Moreover, in order to satisfy the slice isolation constraint and given the maximum allowable number of PRBs allocated to each slice at each gNB, $\overline{R}^k_j$, we formulate the slice isolation constraint using the following inequality:
\begin{equation}
    z^k_{j}W^r_j \leq \overline{R}^k_j x^{k,j^\prime}_j, \ \forall k\in\mathcal{K}, j\in V_{gNB}, j^\prime \in V_{k,gNB}^\prime \label{SIMP:isolation}
\end{equation}
Moreover, for each infrastructure node and link, we have capacity constraints as follows:
\begin{equation}
    \sum_{k\in \mathcal{K}} \sum_{j^\prime\in V^\prime_k} x_{j}^{k,j^\prime} R_{j^\prime}^\nu \chi_k^\nu \leq W_j^\nu, \ \nu\in \mathcal{T}, \forall j\in V_{} \label{SIMP:node-Cap1}
\end{equation}
\begin{equation}
    \sum_{k\in\mathcal{K}} \sum_{p\in \mathcal{P}_k: e\in p}\sum_{e^\prime \in E^\prime_k} y^{k,e^\prime}_p R^{BW}_{e^\prime} \chi_k^{BW} \leq W_e^{BW},\ \forall e \in E
    \label{SIMP:link-Cap1}
\end{equation}
Furthermore, the capacity constraint of gNB $j\in V_{gNB}$ is enforced by the following inequality: 
\begin{equation}
    \sum_{k=1}^K z_{j}^k = x^{k,j^\prime}_j,\ \forall j\in V_{gNB}, j^\prime \in  V_{k,gNB}^\prime \label{SIMP:gNBcap}\\
\end{equation}
The gNB placement constraint is enforced by the following:
\begin{equation}
    x_j^{k, j^\prime} \geq \mathbbm{I}{\{\sum_{i=1}^{U_k} u_{j}^{k,i}\geq1\}},\ \forall k\in \mathcal{K}, j\in V_{gNB}, j^\prime \in V_{k,gNB}^\prime \label{SIMP:gNBplacement2}\\
\end{equation}
Finally, the domain constraints of the second-stage problem are:
\begin{align}
    & z^k_j \in [0, 1],\ \forall k\in \mathcal{K}, j\in V_{gNB}\nonumber\\ 
    & \eta^{k,e^\prime}_p, \eta_p \geq 0,\ \forall k\in \mathcal{K}, e^\prime \in E^\prime_k, p\in \mathcal{P}_k \label{domain2} 
\end{align}

The E2E network slicing problem with demand uncertainty is formulated as a two-stage SMIP presented below:
\begin{align}
    & {minimize} \ \phi_1\mathcal{C}_1(\boldsymbol{x},\boldsymbol{y}) + \phi_2 \ \mathbb{E}_{
    \boldsymbol{\xi}} [min \ \mathcal{C}_2 (\boldsymbol{z})+\epsilon \mathcal{D(\boldsymbol{\eta})}] \label{SMIP:obj} \\
    & \qquad \qquad \qquad \text{s.t.}  \qquad \eqref{SIMP:non-gNB-map1}-\eqref{domain2} \nonumber
\end{align}

The objective \eqref{SMIP:obj} minimizes the summation of the core slice provisioning cost and the expectation of the second stage objective which is the RAN slice provisioning cost. The weights $\phi_1,\phi_2$ determine the balance between the objectives of the first and second stage problems. An interpretation of the \emph{SMNS} is as follows: 
\begin{itemize}
    \item The InP provisions the core network resources for different NSs, represented by the decision variables $\boldsymbol{x}, \boldsymbol{y}$, before the actual value of the random vector 
    \begin{align}
    & \boldsymbol{\xi}=(\mathcal{U}_1, \dots, \mathcal{U}_K, u^{1,1}_{1}, \dots, u^{1,{U}_1}_{1}, \dots, u^{K,\mathcal{U}_K}_{|V_{gNB}|},\nonumber\\ 
    &\chi^\nu_1, \dots, \chi^\nu_K, \nu\in \mathcal{T}\cup\{BW\}) \nonumber
    \end{align} is realized.
    \item After the realization of $\boldsymbol{\xi}$, the RAN resource provisioning and delay decision variables, denoted by $\boldsymbol{z,\eta}$, are determined.
\end{itemize}
The expectation term in \eqref{SMIP:obj} requires an integration over the high-dimensional random vector $\boldsymbol{\xi}$. To tackle this challenge, we use the sample average approximation technique and replace the expectation in \eqref{SMIP:obj} with its sample average approximation (SAA). The samples can be viewed as historical observed data or can be generated using Monte Carlo sampling techniques.
\subsection{Deterministic Equivalent Reformulation}
\label{sec:DEF}
In this section, we introduce the deterministic reformulation of the modeled \emph{SMNS} and then propose a solution based on the SAA. The SAA method is an approach for solving stochastic optimization problems by using Monte Carlo simulation. Suppose that $H$ i.i.d observations of the random variables $u_{j}^{k,i}$, $U_k$, and $\chi_k^\nu$ are available, denoted by $\tilde{u}_{j,h}^{k,i}$,$\Tilde{U}_{k,h}$, $\Tilde{\chi}^\nu_{k,h}$, $h = 1,\dots,H$. For each realization, we define a separate set of second-stage decision variables $z_{j,h}^k, \eta_{p,h}, \eta_{p,h}^{k,e^\prime}$. Thus, we convert \emph{SMNS} to its sampled deterministic equivalent problem, defined as follows:

\begin{align}
&{minimize} \quad \phi_1\mathcal{C}_1(\boldsymbol{x},\boldsymbol{y}) +\phi_2 \sum_{h=1}^H \frac{1}{H} \left( \mathcal{C}_2(\boldsymbol{z}_h) + \epsilon \mathcal{D}(\boldsymbol{\eta}_h)\right)
\label{obj_det} 
\end{align}

\begin{align}
    &{subject \text{ }to:}\quad\quad\quad\quad\quad\quad\eqref{SIMP:non-gNB-map1} - \eqref{SMIP:dom1}\quad\quad\quad\quad\quad\quad\quad\quad\quad\quad\quad\nonumber
\end{align}
\begin{align}
& \sum_{e\in q}\left[ \sum_{k\in\mathcal{K}} \sum_{p\in\mathcal{P}_k:e\in p} \sum_{e^\prime \in E^\prime_k} y^{k,e^\prime}_p\frac{\tilde{\chi}_{k,h}^{BW}R^{BW}_{e^\prime}}{W^{BW}_e}\right] - \eta_{q,h} = 0, \nonumber\\
&\forall q\in \mathcal{P},h\in\mathcal{H} \label{DET-latency-1}
\end{align}
\begin{align}
& \zeta y^{k,e^\prime}_p + \eta_{p,h} - \eta^{k,e^\prime}_{p,h} \leq \zeta, \ \forall p\in\mathcal{P}_k, e^\prime \in E^\prime_k, k\in\mathcal{K}, h\in \mathcal{H}\label{DET-latency-2}
\end{align}
\begin{align}
& \sum_{e^\prime \in p^\prime} \sum_{p\in \mathcal{P}_k} \eta^{k,e^\prime}_{p,h} + \sum_{j^\prime \in p^\prime} \sum_{j\in V} x^{k,j^\prime}_j \tau^{k,j^\prime} \leq d^{UP/CP}_k,\nonumber\\
&\forall p^\prime \in \mathcal{P}^\prime_{k,UP/CP}, k\in \mathcal{K}, h\in \mathcal{H} \label{DET-latency-3}
\end{align}
\begin{align}
& \sum_{k\in \mathcal{K}} \sum_{j^\prime\in V^\prime_k} x_{j}^{k,j^\prime} R_{j^\prime}^{\nu} \tilde{\chi}^\nu_{k,h} \leq W_j^{\nu},\ \forall \nu\in \mathcal{T}, j\in V_{}, h\in\mathcal{H} \label{DET-cap1}
\end{align}
{
\begin{align}
& \sum_{k\in\mathcal{K}} \sum_{p\in \mathcal{P}_k: e\in p}\sum_{e^\prime \in E^\prime_k} y^{k,e^\prime}_p R^{BW}_{e^\prime} \tilde{\chi}^{BW}_{k,h} \leq W_e^{BW},\nonumber\\& \forall e \in E, h\in\mathcal{H} \label{DET-cap2}
\end{align}
}
\begin{align}
&  \sum_{i=1}^{\tilde{U}_{k,h}}\tilde{u}_{j,h}^{k,i}\underline{R}^k_j \leq z^k_{j,h}W^r_j+(1-x^{k,j^\prime}_j)M, \nonumber \\ 
&\forall k\in\mathcal{K}, j\in V_{gNB},j^\prime \in V_{k,gNB}^\prime, h\in \mathcal{H} \label{rate_det1}
\end{align}
\begin{align}
& z^k_{j,h}W^r_j  \leq \overline{R}^k_j x^{k,j^\prime}_j, \ \forall k\in\mathcal{K}, j\in V_{gNB}, j^\prime \in V_{k,gNB}^\prime, h\in \mathcal{H} \label{rate_det2}
\end{align}
\begin{align}
&  \sum_{k=1}^K z_{j,h}^k \leq 1, \ \forall j\in V_{gNB}, h\in\mathcal{H} \label{cap_ran_det}
\end{align}
{
\begin{align}
&  x_j^{k, j^\prime} \geq \mathbbm{1}{\{\sum_{i=1}^{\Tilde{U}_{k,h}} \tilde{u}_{j,h}^{k,i}\geq1\}},\nonumber\\& \forall k\in \mathcal{K}, j\in V_{gNB}, j^\prime \in V_{k,gNB}^\prime, h\in\mathcal{H} \label{placement_det2}
\end{align}}
\begin{align}
& z^k_{j,h} \in [0,1], \eta_{p,h}, \eta_{p,h}^{k,e^\prime}\geq 0, \ \forall k\in \mathcal{K}, j\in V_{gNB}, h\in\mathcal{H} \label{DET-domain_det}
\end{align}
We refer to the above problem as $\emph{DET-SMNS}(\mathcal{H})$. The first term in the objective function \eqref{obj_det} is the same as the first term of the objective function \eqref{SMIP:obj}. The second term is the weighted average of the second stage objective function over the $H$ realized samples of the random variables $\chi_k$, $u^k_{i,j}$ and $U_k$. Constraints \eqref{DET-latency-1}-\eqref{DET-domain_det} replace their counterparts.

It is shown in \cite{shapiro1996simulation} that as the sample size $H$ increases and under certain mild conditions as discussed in \cite{shapiro1996simulation}, the solution to $\emph{DET-SMNS}(\mathcal{H})$ converges to that of the original \emph{SMNS} problem. In the next section, we present a practical algorithm for the resource provisioning of network slicing incorporating the \emph{DET-SMNS} model.

\section{Two-Timescale Resource Provisioning}
\label{sec:solution}
We propose a two-timescale resource allocation scheme that utilizes the resource provisioning methodology discussed in Sections \ref{sec:sys} and \ref{sec:formulation_thesis}. In this approach, the E2E resource allocation problem of NSs is addressed through two long and short time slots. We assume that the lifecycle of an NS is divided into a number of long time slots (macro-slots) indexed by $T$. During each macro-slot, the demand distribution for NSs is known and fixed. Let $\mathcal{F}^T = \{\mathcal{F}^T_1,\dots,\mathcal{F}^T_K\}$, where $\mathcal{F}_k(.,.):\mathbb{R}^2 \xrightarrow{} [0,1]$ denote the spatial density function of the $k$th slice demand across the considered geographical area in the macro-slot $T$. Given $\mathcal{F}^T$, we solve the resource provisioning problem proposed in Section \ref{sec:DEF}, $\emph{DET-SMNS}(\mathcal{H})$.

We further divide each macro-slot into $N_T$ short time slots (micro-slots). At each micro-slot $t$, the actual demand is observed that is shown by $\hat{\boldsymbol{\xi}}=(\hat{\mathcal{U}}_k, \hat{u}^{k,i}_{j}, \hat{\chi}^\nu_k, k\in\mathcal{K},i\in \hat{\mathcal{U}}_k, j\in V_{gNB},\nu\in\mathcal{T}\cup\{BW\})$. We define the decision variable $\sigma^k_j, k\in \mathcal{K}, j\in V_{gNB}$ as the fraction of unsupported demand requested for slice $k$ in gNB $j$. Given the observed demand at each micro-slot, and the solution to the first-stage problem ($\boldsymbol{x}, \boldsymbol{y}$), we adjust the RAN resource provisioning decision by solving the following linear program (LP):
\begin{align}
    & {minimize}\quad  \sum_{k\in\mathcal{K}}\sum_{j\in V_{gNB}} \theta (\mathcal{C}_2(\boldsymbol{z})) +  (1-\theta)\sigma^k_j \label{real:obj}
\end{align}
\begin{align}
    & (1-\sigma^k_j)(\sum_{i=1}^{\hat{{U}}_k} \hat{u}^{k,i}_{j}) \underline{R}^k_j \leq z^k_j W^r_j + (1-x^{k,j^{\prime}}_{j})M, \nonumber\\ 
    &\forall k\in \mathcal{K}, j\in V_{gNB}, j^\prime \in V_{k,gNB}^\prime \label{real:one}
\end{align}
\begin{align}
    & \frac{\sum_{j\in V_{gNB}}(1-\sigma^k_j) \sum_{i=1}^{\hat{U}_k}\hat{u}^{k,i}_j}{\hat{U}_k}\hat{\chi}^\nu_k \leq \chi^\nu_{k,prov}, \nonumber\\ 
    &\forall k\in\mathcal{K}, \nu\in \mathcal{T}\cup \{BW\} \label{real:two}
\end{align}
\begin{align}
    & z^k_j W^r_j \leq \overline{R}^k_j x^{k,j^\prime}_j, \ \forall k\in \mathcal{K}, j\in V_{gNB}, j^\prime \in V^\prime_{k,gNB}\label{real:three}
\end{align}
\begin{align}
    & \sum_{k\in\mathcal{K}}z^k_j\leq 1,\ \forall j \in V_{gNB} \label{real:four}
\end{align}
\begin{align}
    & z^k_j, \sigma_j^k \in [0,1], \ \forall k\in \mathcal{K}, j\in V_{gNB} \label{real:five}
\end{align}
We refer to this LP as the radio network slicing resource allocation, denoted by $\emph{RNSR}(\boldsymbol{x},\boldsymbol{y}, \hat{\boldsymbol{\xi}} )$.

The objective of \eqref{real:obj} minimizes the weighted summation of RAN resource allocation cost and total unsupported traffic. The weight parameter $\theta \in [0,1]$ is used to balance between the two objective terms. Constraint \eqref{real:one} ensures that the radio resource reservation for an NS is greater than the supported traffic on each gNB. $M$ is a big positive real value used for the big-M method, enforcing constraint \eqref{real:one} for a tuple $(j', k, j)$ only if $x^{k,j'}_j = 1$, i.e. a RAN VNF of slice $k$ is instantiated on gNB $j$. Constraint \eqref{real:two} guarantees the availability of core network resources to support the admitted traffic of each slice. The slice isolation constraint is enforced by \eqref{real:three}. Constraint \eqref{real:four} ensures the capacity constraint for each gNB, and the domain constraints are expressed by \eqref{real:five}. Algorithm \ref{algo}  illustrates our proposed two-timescale resource provisioning scheme.

\begin{algorithm}[t]
 \caption{Two-timescale Resource Provisioning}
 \label{algo}
 \begin{algorithmic}[1]
 \Statex \textbf{Input: } $G$, $G^\prime_k, \mathcal{U}_k, k\in\mathcal{K}$
 \Statex \textbf{Output: } $\boldsymbol{x}$, $\boldsymbol{y}$, $\boldsymbol{z}, \sigma$
 \For{$T=1,\dots$}
 \State Find $\mathcal{F}^T$ using historical demand distribution data.
 \State Take $H$ sample of $\mathcal{F}^T$, denoted by the set $\mathcal{H}$.
 \State $(\boldsymbol{x, y}) \xleftarrow{}$ Solve $\emph{DET-SMNS}(\mathcal{H})$.
 \For{$t=1,\dots,N_T$}
 \State Observe $\hat{\boldsymbol{\xi}}$.
 \State $(\boldsymbol{z, \sigma})\xleftarrow{}$ Solve $\emph{RNSR}(\boldsymbol{x},\boldsymbol{y},\hat{\boldsymbol{\xi}})$
 \EndFor
 \EndFor
 \end{algorithmic} 
 \end{algorithm}

\section{Performance Evaluation}
\label{sec:evaluation}

In this section, we present the evaluation of the proposed two-timescale methodology through numerical simulations. We first describe the simulation setup and parameters and then proceed to the numerical results. 
\begin{figure}[!t]
    \centering
    \includegraphics[width=0.45\textwidth]{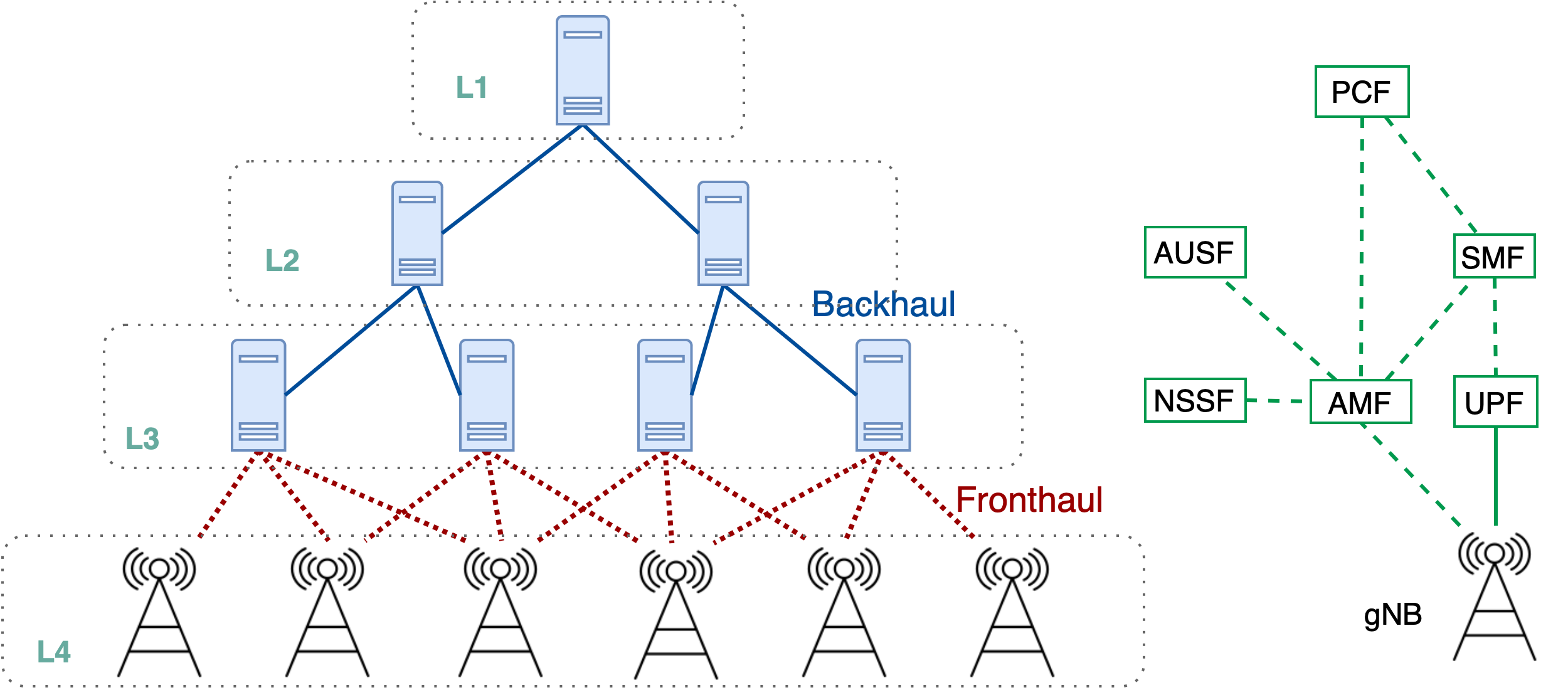}
    \caption{Example substrate (left) and NS (right)}
    \label{fig:ns-combined}
\end{figure}
\begin{table}[t]
\centering
\caption{System Model Parameters}
\begin{center}
\scalebox{1}{
\begin{tabular}{|c|c|c|c|}
\hline
\hline
Slice Type & $d_k^{UP}$/$d_k^{CP}$ & mean value of $\underline{R}^k_j$ & $\overline{R}^k_j$\\
\hline
\hline
eMBB (k=1) & 100/20 ms & 50 & 0.6\\
\hline
URLLC (k=2) & 25/5 ms & 10 & 0.2\\
\hline
mMTC (k=3) & 300/60 ms & 5 & 0.2\\
\hline
\end{tabular}}
\end{center}
\label{chapter2:sim_parameters}
\end{table}
\begin{figure*}[t]
\centering

\begin{subfigure}{0.32\textwidth}
  \includegraphics[width=\textwidth]{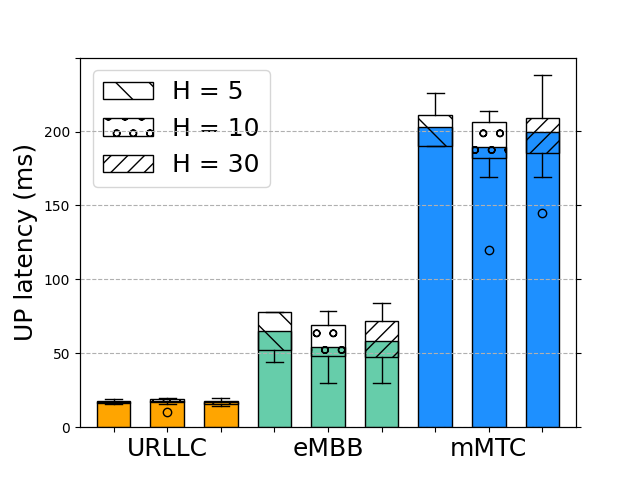}
  \caption{UP latency for different network slices}
    \label{chapter2:delay-up}
\end{subfigure}
\begin{subfigure}{0.32\textwidth}
  \includegraphics[width=\textwidth]{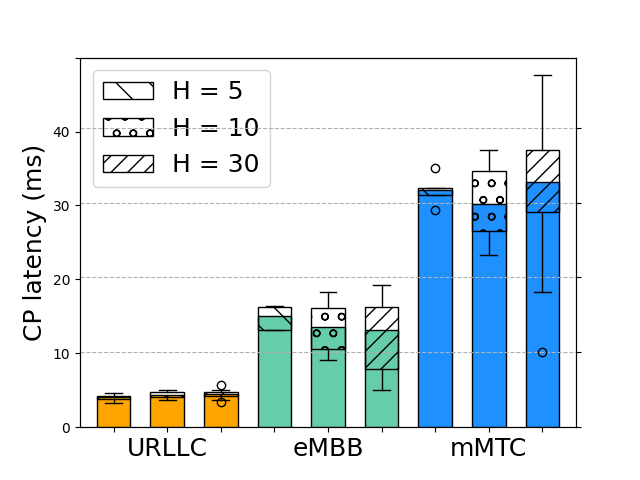}
  \caption{CP latency for different network slices}
    \label{chapter2:delay-cp}
\end{subfigure}
\caption{UP and CP latency for URLLC, eMBB, and mMTC slices}
\end{figure*}
\begin{figure*}[t]
\centering
\begin{subfigure}{0.32\textwidth}  \includegraphics[width=\textwidth]{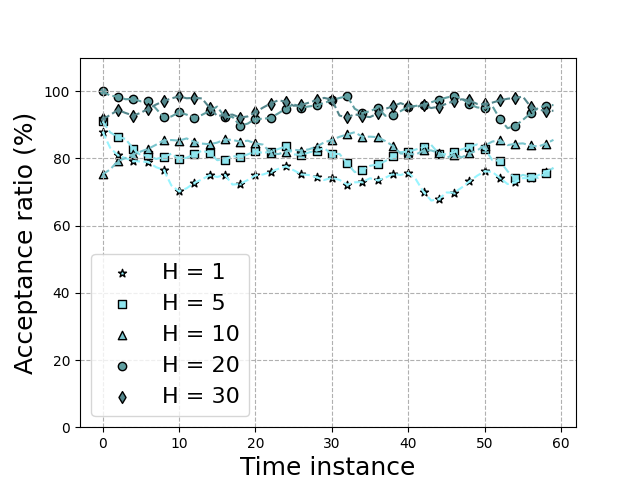}
  \caption{eMBB slice}
    \label{chapter2:acc-embb}
\end{subfigure}
\begin{subfigure}{0.32\textwidth}
  \includegraphics[width=\textwidth]{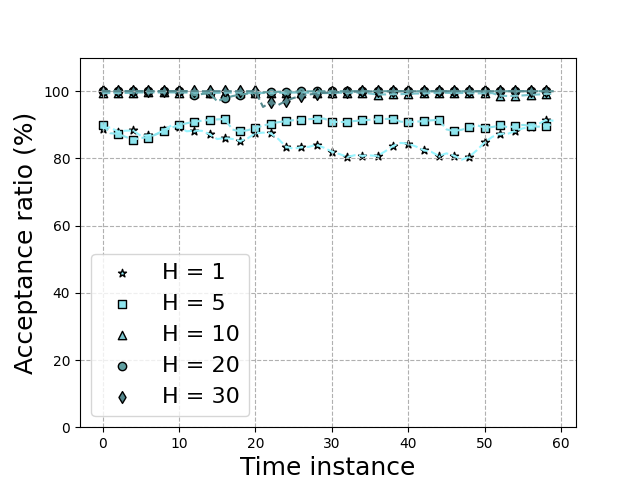}
  \caption{URLLC slice}
    \label{chapter2:acc-urllc}
\end{subfigure}
\begin{subfigure}{0.32\textwidth}
  \includegraphics[width=\textwidth]{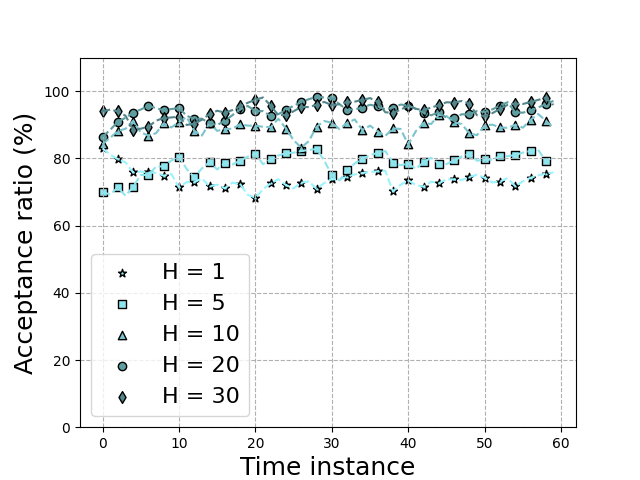}
  \caption{mMTC slice}
    \label{chapter2:acc-mmtc}
\end{subfigure}
\caption{Average request acceptance ratio for URLLC, eMBB, and mMTC slices}
\end{figure*}

\subsection{Simulation Setup}
We consider a mobile network consisting of seven general-purpose server nodes at four levels of hierarchy and $6$ gNB nodes connected to the third-level servers as illustrated in Fig. \ref{fig:ns-combined}. L1, L2, L3, and L4 servers have CPU (cycle/s), storage (GB), and RAM (GB) capacities of $(72, 144, 288)$, $(36, 72, 144)$, $(18, 36, 72)$, and $(6, 12, 24)$, respectively. The backhaul links have a capacity of $2$Gbps while the fronthaul links can support $1$Gbps of traffic. For the sake of simplicity, the resource scaling factor $\chi^\nu_k$ is assumed to be the same for all resource types $\nu \in \mathcal{T}$ and is equal to the number of users requesting slice $k$. Each VNF requires $(0.1\chi_k,0.2\chi_k,0.4\chi_k)$ units of CPU, STO, and RAM resources. The simulation environment is implemented in Java and we use the CPLEX commercial solver for solving the $\emph{DET-SMNS}$ model using the branch-and-bound method. The $\emph{RNSR}$ problem is also solved by CPLEX using the Simplex method. All experiments are carried out on an Intel Xeon processor at 2.3 GHz with 8GB memory.

We assume that the UEs are distributed in a square area with the dimension of $10\ Km\times10\ Km$. The demand density functions $\mathcal{F}^T$ are obtained by applying bivariate kernel density estimation (KDE) to the synthetic data generated for each slice demand in the considered geographical area. Due to the unavailability of spatially distributed datasets for the demand of NSs, we generate synthetic data for the UE locations using Matérn cluster point process \cite{cluster}. In this process, a number of parent points (UEs) are first generated using a Poisson point process and for each parent point (representing a cluster center), another Poisson point process is used to generate the UE distribution within a radius of the parent point. The cluster-based UE distribution represents the hot spot area for different services supported through network slices. 

We assume that the demand density functions $(\mathcal{F}^T)$ change hourly, i.e. the value of the macro-slot is one hour. Moreover, the duration of the micro-slot is set to one minute, i.e. $N_T = 60$. We consider three slices, namely, eMBB, URLLC, and mMTC with the SFC graph as depicted in Fig.~\ref{fig:ns-combined}. The simulation parameters of different NS types are given in Table \ref{chapter2:sim_parameters}.
It is worth noting that the duration of the macro-slot and micro-slot can be adjusted with respect to the slice type and the dynamicity of the considered scenario, e.g. UEs mobility parameters such as velocity, channel condition variations, etc. We use the $k$-shortest path algorithm to construct the set $\mathcal{P}( i \rightarrow j ) \forall i,j \in V$  with $k=3$. 

\subsection{Numerical Results}
Figure 4 illustrates the UP and CP latency for the URLLC, eMBB, and mMTC slices. We consider the results of $5$ and depict both the average latency and the profile of latency values for different NSs, in the cases of $H=5,10,30$. We observe that in all cases, the obtained UP and CP latencies are below the corresponding maximum tolerable latency given in Table \ref{chapter2:sim_parameters}. Thus, the proposed two-time-scale resource provisioning algorithm provides solutions that meet the NSs' QoS requirement in terms of E2E UP and CP latency. 

In Fig. 5, the average acceptance ratio of different slices averaged for the duration of $60$ micro-slots is illustrated. The percentage of accepted requests is obtained from the supported traffic solution of the \emph{RNSR} problem, and averaged over different gNBs, i.e.:
$$\text{Avg. acceptance ratio of slice $k$} = 100*\sum_{j\in V_{gNB}}\sigma^k_j/|V_{gNB}|$$  Figure \ref{chapter2:acc-embb}, \ref{chapter2:acc-urllc}, and \ref{chapter2:acc-mmtc} denote the average acceptance ratio for the eMBB, URLLC, and mMTC slices, respectively. In this experiment, we change the number of realizations from $1$ to $30$. It is observed that as the value of $H$ increases (more realizations are considered as an input to $\emph{DET-SMNS}(\mathcal{H)}$), the acceptance ratio enhances for all slices. This is due to the fact that as the number of realizations increases, the solution of the $\emph{DET-SMNS}(\mathcal{H)}$ problem is a better approximation of the original \emph{SMNS} problem modeled as a SMIP as explained in Section \ref{sec:SMNS}.

\section{Conclusion}
\label{sec:conclusion}
In this paper, we study the resource provisioning problem for end-to-end (E2E) network slicing under demand uncertainty. We modeled the problem as a two-stage stochastic mixed integer program (SMIP) aiming at minimizing the total resource cost over the core network (CN) and the radio access network (RAN). We develop a practical two-time-scale resource provisioning algorithm based on our SMIP model. The proposed algorithm operates in two phases with long and short time scales corresponding to the E2E resource provisioning followed by RAN resource adjustments.  We show by means of numerical simulations that the proposed algorithm successfully addresses the resource provisioning problem for network slicing under demand uncertainty. 
\bibliographystyle{IEEEtran}
\bibliography{bib}
\end{document}